\documentclass[journal]{IEEEtran}
\usepackage{amsmath,amssymb}
\usepackage{graphicx,graphics,color,psfrag}
\usepackage{cite,balance}
\usepackage{caption}
\allowdisplaybreaks
\usepackage{algorithm}
\usepackage{algorithmic}
\usepackage{accents}
\usepackage{amsthm}
\usepackage{bm}
\usepackage{url}
\usepackage[english]{babel}
\usepackage{multirow}
\usepackage{enumerate}
\usepackage{cases}
\usepackage{stfloats}
\usepackage{dsfont}
\usepackage{color,soul}
\usepackage{amsfonts}
\usepackage{graphicx}
\usepackage[caption=false,font=footnotesize]{subfig}
\usepackage{cite,graphicx,amsmath,amssymb}
\usepackage{fancyhdr}
\usepackage{hhline}
\usepackage{graphicx,graphics}
\usepackage{array,color}
\usepackage{mathtools}
\usepackage{amsmath}
\usepackage{hyperref}
\usepackage{makecell}

\usepackage{adjustbox}  

\makeatletter

\newcommand{\Rmnum}[1]{\expandafter\@slowromancap\romannumeral #1@}
\makeatother

\newtheorem{proposition}{\emph{\underline{Proposition}}}

\newtheorem{remark}{\bf \emph{\underline{Remark}}}

\def\({\left(}
\def\){\right)}

\def\b0{{\mathbf{0}}}

\newcommand{\nn}{\nonumber}

\begin{document}
\captionsetup[figure]{name={Fig.}}

\makeatletter
  \renewcommand\p@subfigure{}                     
  \renewcommand\thesubfigure{\alph{subfigure}}  
\makeatother

\everymath{%
  \thinmuskip=0.5mu   
  \medmuskip=0.5mu    
  \thickmuskip=0.5mu  
}

\title{\huge 
Two-stage Multi-beam Training for Multiuser Millimeter-Wave Communications } 
\author{Weijia Wang, Changsheng You, \IEEEmembership{Member, IEEE}, Xiaodan Shao , \IEEEmembership{Member, IEEE}, and Rui Zhang, \IEEEmembership{Fellow, IEEE}

\thanks{Weijia Wang is with School of Science and Engineering, The Chinese University of Hong Kong, Shenzhen, Guangdong 518172, China (e-mail: weijiawang@link.cuhk.edu.cn).

Changsheng You is with the Department of Electronic and Electrical Engineering, Southern University of Science and Technology (SUSTech), Shenzhen 518055, China (e-mail: youcs@sustech.edu.cn).

 Xiaodan Shao is with the Department of Electrical and Computer Engi
neering, University of Waterloo, Waterloo, ON N2L 3G1, Canada (email:
 x6shao@uwaterloo.ca).

Rui Zhang is with the Department of Electrical and Computer Engineering, National University of Singapore, Singapore 117583 (e-mail: elezhang@nus.edu.sg).

\textit{Corresponding author: Changsheng You and Xiaodan Shao.}
}}
\maketitle

\begin{abstract}
In this letter, we study an efficient multi-beam training method for multiuser millimeter-wave communication systems. Unlike the conventional single-beam training method that relies on exhaustive search, multi-beam training design faces a key challenge in balancing the trade-off between beam training overhead and success beam-identification rate, exacerbated by severe inter-beam interference. To tackle this challenge, we propose a new \emph{two-stage} multi-beam training method with two distinct multi-beam patterns to enable fast and accurate user angle identification. Specifically, in the first stage, the antenna array is divided into sparse subarrays to generate multiple beams (with high array gains), for identifying candidate user angles. In the second stage, the array is redivided into dense subarrays to generate flexibly steered wide beams, for which a cross-validation method is employed to effectively resolve the remaining angular ambiguity in the first stage. Last, numerical results demonstrate that the proposed method significantly improves the success beam-identification rate compared to existing multi-beam training methods, while retaining or even reducing the required beam training overhead.
\end{abstract}
\begin{IEEEkeywords}
Millimeter wave communication, codebook design, multi-beam training.
\end{IEEEkeywords}

\vspace{-6pt}
\section{Introduction}
\vspace{-3pt}
Millimeter-wave (mmWave) communication is a promising technology to boost network capacity of next-generation wireless systems \cite{Channel_Estimation_2021,el2014spatially}, by providing abundant spectrum resources. In particular, to compensate for the severe path loss in high-frequency bands, a large number of antennas need to be deployed at the base station (BS) to provide high beamforming gains with pencil-like beams \cite{Beam_Alignment_2023}. 

To reap the prominent beamforming gain of large-scale antenna array, beam training is critical for establishing high signal-to-noise ratio (SNR) links \cite{Communicating_With_2021,Near-field_2024,Learning_Site}. However, the conventional single-beam exhaustive search method incurs beam training overhead due to the narrow beams generated by large-scale antenna array\cite{Multiuser_Millimeter}. To address this issue, a hierarchical codebook-based beam training method was proposed in \cite{Hierarchical_Codebook_2020}, where the user angles are progressively resolved by firstly employing wide beams and then narrow beams. Alternatively, multi-beam training is another efficient approach to reduce beam training overhead, by steering multiple beams toward different angles simultaneously. For instance, the authors in \cite{you2020fast} proposed a multi-beam training method by partitioning the antenna array into multiple subarrays, and then employing the cross-validation method to enable fast beam identification. However, this method may suffer a low identification rate when the number of beams is large. Additionally, by exploiting the grating lobes of the uniform sparse array, a two-phase beam training strategy was proposed in \cite{zhou_multi-beam_2024}, which performs multi-beam sweeping in the first phase to determine candidate angles by using the antenna sparse-activation method, followed by the second phase to identify user angle via single-beam sweeping. Nevertheless, the beam-identification rate of this method is not high in the low-SNR regime due to the power loss in sparse antenna-activation.

To address the above issues, we propose in this letter a new \emph{two-stage} multi-beam training method for multiuser mmWave communication systems to enhance success beam-identification rate while retaining low beam training overhead. Specifically, in the first stage, we divide the entire array into multiple sparse subarrays to generate uniformly spaced narrow beams, hence accurately identifying candidate user angles. Then, in the second stage, we redivide the antenna array into multiple dense subarrays to form flexibly steered wide beams, which enable rapid beam sweeping over the entire space to resolve the remaining angular ambiguity in the first stage. Last, numerical results demonstrate that our proposed method achieves a superior trade-off between beam training overhead and success beam-identification rate compared to existing multi-beam training methods.

\vspace{-6pt}

\begin{figure}[!t]
    \centering
    \includegraphics[width=6cm]{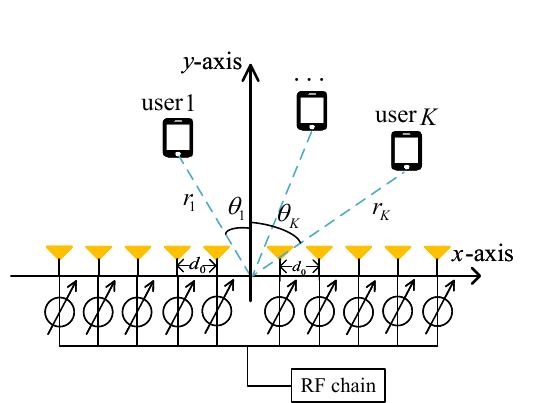}
    \caption{A narrow-band far-field multiuser beam training system.}
    \label{fig:sytem_model}
    \vspace{-16pt}
\end{figure}

\vspace{-6pt}
 
\section{System Model}
\vspace{-3pt}
As shown in Fig.~\ref{fig:sytem_model}, we consider downlink beam training for a multiuser communication system, where a BS equipped with an $N$ antenna uniform linear array (ULA) communicates with $K$ single-antenna users. 

\underline{\bf Channel model:}
The ULA is assumed to be placed along the $x$-axis, for which the position of the $n$-th antenna is given by $(d_{n}, 0)$ with $d_{n}=\frac{2n-N-1}{2}d_{0}$, $n=1,2,\ldots,N$, and $d_0=\frac{\lambda}{2}$ representing the half-wavelength. The served $K$ users are denoted by a set $\mathcal{K}=\{1,2,\ldots,K\}$, which are distributed on the $x$-$y$ plane. For each user $k$, its physical angle-of-departure (AoD) from the BS is denoted as $\theta_k \in \left[ -\frac{\pi}{2}, \frac{\pi}{2} \right]$, and the corresponding spatial angle is $\Omega_{k}=\sin\theta_{k}$.

We consider multiuser mmWave communication systems, for which the line-of-sight (LoS) path dominates and the non-LoS (NLoS) paths have negligible powers. Let $\mathbf{v}(\Omega, N)\triangleq \left[1,e^{-\jmath  \phi},\ldots,e^{-\jmath  (N-1) \phi}\right]^{T}$ denote the channel steering vector, where $\phi = \frac{2\pi d_0 \Omega}{\lambda}$ represents the phase difference between the signals received at two adjacent antennas, and $\jmath$ denotes the imaginary unit. As such, the LoS channel from the BS to user $k$, denoted by $\mathbf{h}^{H}_{k} \in \mathbb{C}^{1 \times N}$, can be modeled as \cite{el2014spatially}
 \begin{equation}\label{channel}
     \mathbf{h}_{k}^{H} = g_{k} \mathbf{v}^{H}( \Omega_{k},N), \text{ }  \forall k \in \mathcal{K},
 \end{equation}
 where $g_{k}=\beta \left(\frac{r_{0}}{r_{k}}\right)^{\alpha}e^{-\frac{\jmath2\pi r_{k}}{\lambda}}$ is the complex-valued path gain of the BS-user link, with $\beta$ and $r_{k}$ respectively denoting the reference channel gain at a distance of $r_{0}=1\text{ meter (m)}$ and the distance between the BS and user $k$, and $\alpha$ is the path-loss exponent. 
 
 \underline{\bf Signal model:} 
 Based on \eqref{channel}, the received signal at each user $k$ is given by
 \begin{equation} \label{signal}
     y_k =  \mathbf{h}_k^H \mathbf{f} s + z_k=g_{k}\mathbf{v}_{k}^{H}(\Omega_{k},N)\mathbf{f}s+z_{k}, \text{ } \forall k \in \mathcal{K},
 \end{equation}
 where $s$ is the transmitted (training or information) signal by the BS with average power $P$. $\mathbf{f}\in \mathbb{C}^{N\times1}$ represents the beamforming vector (for training or data transmission), and $z_k$ is the received additive white Gaussian noise (AWGN) at user $k$ with average power $\sigma^2$.
 
We consider a two-phase transmission protocol. Specifically, in the first phase, the BS employs a dedicated beam-training codebook (denoted by $\mathbf{W}$ which is to be designed) for estimating spatial angle $\hat{\Omega}_{k}$ for each user $k$. Then, in the second phase, the BS selects a directional beam in a predefined codebook (denoted by $\mathbf{U}\triangleq\{\mathbf{u}_{1},\mathbf{u}_{2},\ldots,\mathbf{u}_{N}\}$) for its data transmission. Note that for far-field communications, discrete Fourier transform (DFT) codebook has been widely used for data transmission, which is defined as 
  \begin{align}
        \mathbf{u}_n \triangleq \frac{1}{\sqrt{N}} \left[ 1, e^{\jmath \pi  \Omega_{n}}, \ldots, e^{\jmath \pi (N-1) \Omega_{n}} \right]^T,\text{ } \forall n\in\{1,2,\ldots,N\},
  \end{align}
  where $\Omega_{n}=-1+(2n-1)/N$. For each user $k$, to maximize the beamforming gain in data transmission, its optimal beam index that achieves the best beam alignment with channel steering vector can be easily obtained as \footnote{\vspace{-2pt} During the beam training phase, a higher received SNR generally leads to more accurate angle estimation. Such improved estimation directly contributes to higher beamforming gain in the data transmission phase.}
\begin{equation}\label{optimal}
    \hat{n}_k = \underset{n \in \{1,2,\ldots,N\}}{\arg \max} | \mathbf{v}_k^H(\hat{\Omega}_{k},N) \mathbf{u}_n |^2, \text{ } \forall k \in \mathcal{K}. 
\end{equation} 

In the following, we focus on designing both efficient beam training codebook $\mathbf{W}$ as well as multi-beam training method with it.

\section{Benchmark Schemes}
\vspace{-3pt}
Note that one straightforward method to find the optimal beam index for each user is by applying the single-beam training based on exhaustive search, which, however, incurs prohibitively high beam training overhead when the number of antennas is large. In this section, we introduce two typical multi-beam training methods designed in the existing literature to address this issue and point out their main limitations.
\vspace{-8pt}

\subsection{Multi-Beam Training Based on Dense Subarrays}
Multi-beam training based on dense subarrays is an efficient method to reduce the beam training overhead of the exhaustive search \cite{you2020fast}. Specifically, to generate $Q$ distinct beams towards different angles, we can divide the entire antenna array into $Q$ subarrays, which steer multi-beams towards different angles flexibly. Leveraging this property, the BS performs multiple beam sweeping over time by employing a dedicated multi-beam codebook, after which each user employs the cross-validation criteria to determine its best beam index based on received signal powers. For the $K$-user case, this scheme requires a total number of $ T^{\rm DS} = \frac{N}{Q} \big(1+ \frac{\log_2 Q}{2}\big)$ training symbols \cite{you2020fast}. However, there may exist severe inter-beam interference when a large number of beams are formed concurrently, resulting in degraded success beam-identification rate.

\vspace{-6pt}
\subsection{Multi-Beam Training Based on Antenna Sparse-Activation}

An alternative multi-beam training method is by exploiting the grating lobes of uniform sparse arrays for generating multiple beams simultaneously\cite{zhou_multi-beam_2024}, which is termed as antenna sparse-activation-based multi-beam training. Specifically, to generate $Q$ beams, we can activate one antenna for every $Q$ antennas (i.e., $Q$ represents the sparsity of activated antennas). For beam training, in the first phase, the BS only needs to perform beam sweeping in an angular \emph{subspace} $ [-1,-1+\frac{2}{Q}]$ and each user feeds back its candidate beam indices that yield sufficiently high received signal powers\cite{zhou_multi-beam_2024}. Then, to resolve the angular ambiguity, single-beam sweeping is performed in the second phase to determine the true user angle among candidate angles. For the $K$-user case, this method requires a total number of $ T^{\rm ASA}=\frac{(N-1)}{Q}+KQ$ training symbols. However, this antenna sparse-activation method inevitably suffers degraded array gain due to antenna sparse activation, hence reducing the success beam-identification rate especially in the low-SNR regime.

\vspace{-6pt}
\section{Proposed Two-Stage Multi-Beam Training Method}
\vspace{-3pt}
In this section, we propose a two-stage multi-beam training method that improves the beam-identification rate with low training overhead.

\begin{figure}[!t]
  \centering
  \setlength{\tabcolsep}{4pt}
  \begin{tabular}{c|c}
    {\renewcommand\thesubfigure{a}
    \subfloat[Sparse-subarray multi‐beam pattern with $N=32$, $M^{(\rm I)}=4$.]{%
      \includegraphics[width=0.4\linewidth]{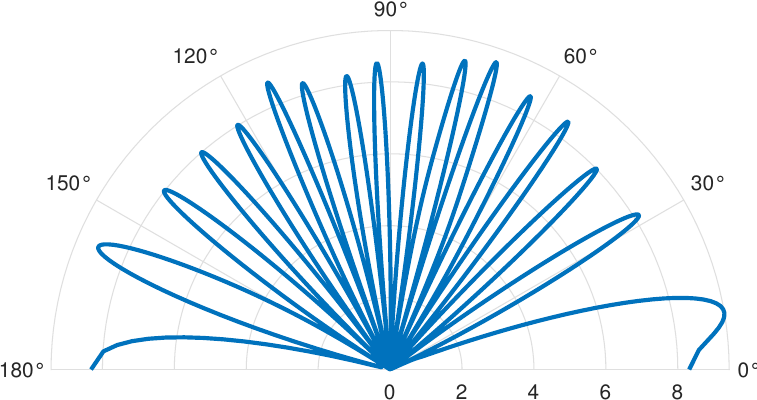}
      \label{fig2.1}
    }
    }&
    {\renewcommand\thesubfigure{c}%
    \subfloat[Illustration of first stage beam training with $N=16$, $M^{(\rm I)}=2$.]{%
      \includegraphics[width=0.5\linewidth]{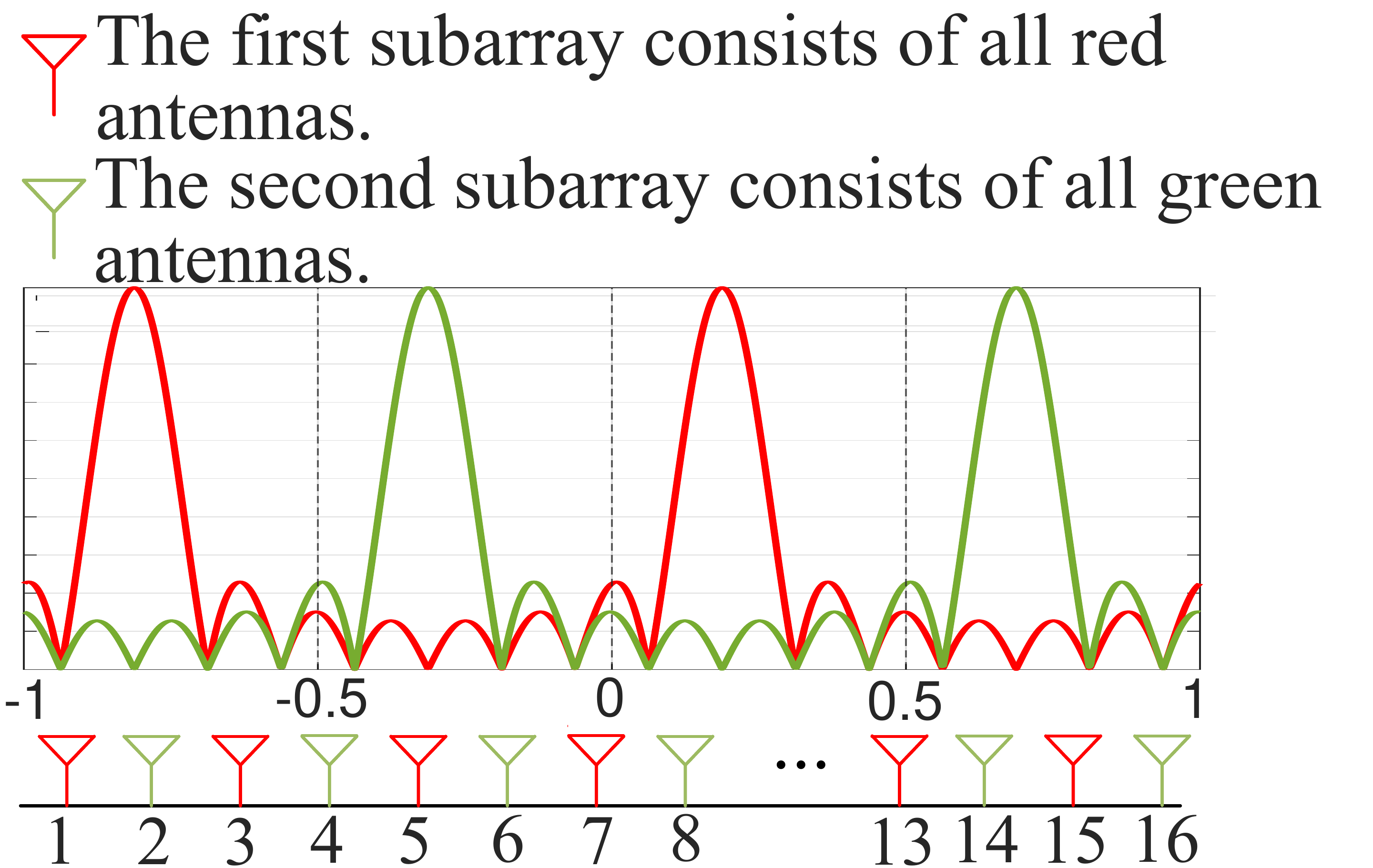}%
      \label{fig2.3}
    } 
    }

    \\
    \cline{1-2}
    {\renewcommand\thesubfigure{b}
    \subfloat[Dense-subarray multi‐beam pattern with $N=32$, $M^{(\rm I \kern-0.1em I)}=4$.]{%
      \includegraphics[width=0.4\linewidth]{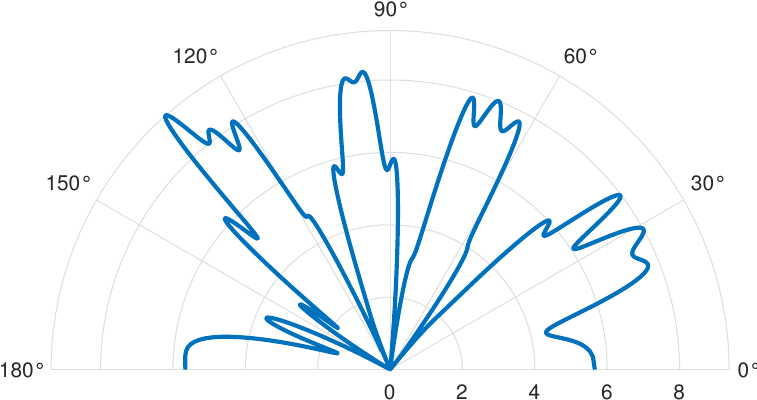}%
      \label{fig2.2}
    }
    }&
    {\renewcommand\thesubfigure{d}
    \subfloat[Illustration of second stage beam training with $N=16$, $M^{(\rm I \kern-0.1em I)}=2$.]{%
      \includegraphics[width=0.5\linewidth]{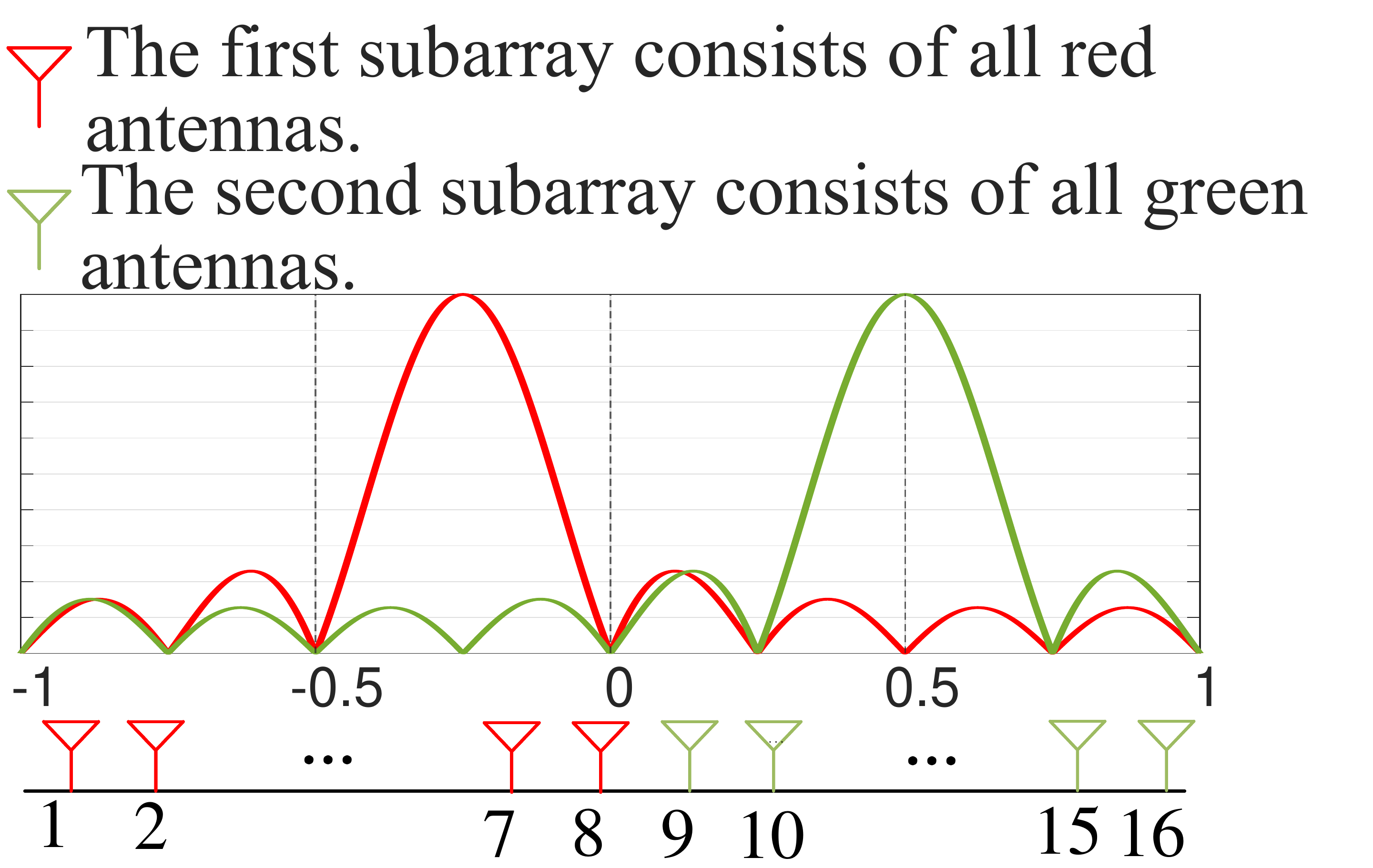}%
      \label{fig2.4}
    } 
    }
    \\ 
  \end{tabular}
  \caption{Illustration of two-stage multi-beam training.}
  \label{fig:2x3grid_with_lines}
  \vspace{-6pt}
\end{figure}

In Stage 1, we divide all antennas into multiple sparse subarrays (as shown in Fig. 2(\ref{fig2.3})) to generate multiple grating-lobe beams for identifying candidate user angles and avoid the power loss in conventional sparse-activation. In Stage 2, we redivide the array into dense subarrays with half-wavelength spacing (as shown in Fig. 2(\ref{fig2.4})) and employ cross-validation to resolve the angular ambiguity with low multiuser training overhead.

We next detail the two-stage multi-beam training design.

 \textbf{First stage (estimation of candidate angles):}  We first design a new sparse-subarray based multi-beam codebook and the associated beam sweeping method.
    
     \emph{1)  Sparse-subarray based codebook:} \emph{1) Sparse-subarray based codebook:} Unlike the conventional sparse-activation method, we design a multi-beam codebook using all antennas (instead of a portion of antennas), which preserves the narrow beamwidth of \cite{zhou_multi-beam_2024} but achieves higher array gain. To this end, we first divide the entire array into $M^{(\mathrm{I})}$ sparse subarrays as shown in Fig. 2(c), each consisting of $N/M^{(\rm I)}$ antennas with an inter-antenna spacing of $\lambda M^{(\rm I)}/2$. As such, for each subarray $m$, its steered spatial angle $\overline{\Omega}_{m},\text{ } m\in\{1,2,\ldots,M^{(\rm I)}\}$ can be flexibly controlled, whose corresponding beam steering vector $\overline{\mathbf{w}}_{m}$ is given by $ \overline{\mathbf{w}}_{m}=e^{-\jmath \pi (m-1)\overline{\Omega}_{m} } \left[1,e^{-\jmath \pi M^{(\rm I)} \overline{\Omega}_{m} },\ldots,e^{-\jmath \pi (N/M^{(\rm I)}-1)M^{(\rm I)} \overline{\Omega}_{m} }\right]^{T}$. The full array steering vector is
     \begin{equation}
     \label{codeword1}
         \mathbf{w}^{(\mathrm{I})} = \operatorname{vec}\left( [\overline{\mathbf{w}}_1, \overline{\mathbf{w}}_2, \ldots, \overline{\mathbf{w}}_{M^{(\mathrm{I})}}] ^{T}\right),
     \end{equation}
      where $\operatorname{vec}(\cdot)$ represents vectorization. This interleaves the subarray steering vectors into the full array. Based on the above, we design the sparse-subarray based multi-beam pattern as follows. 
    
    \begin{proposition}[Multi-beam pattern of sparse subarrays]\label{prop1}
        \emph{For the each sparse subarray $m$, by setting its steered angle as $ \overline{\Omega}_{m} ,\text{ } \forall m\in\{1,2,\ldots,M^{(\rm I)}\}$, the beam codeword $\mathbf{w}^{(\rm I)}$ in (\ref{codeword1}) generates a multi-beam pattern, with $Q^{(\rm I)}=(M^{(\rm I)})^{2}$ directional beams steered towards the following angles,
            \begin{align}\label{angleset}
             \overline\Omega_m + \tfrac{2\,(j-1)}{M^{(\rm I)}},
            \text{ } m=1,\dots,M^{(\rm I)},\text{ } j=1,\dots,M^{(\rm I)}.
            \end{align}
         Specifically, for each sparse subarray with $N/M^{(\rm I)}$ antennas and an inter-antenna spacing of $\lambda M^{(\rm I)}/2$, it generates $M^{(\rm I)}$ beams steered towards $\{\overline{\Omega}_{m}, \overline{\Omega}_{m}+2/M^{(\rm I)},\ldots, \overline{\Omega}_{m}+2(M^{(\rm I)}-1)/M^{(\rm I)}\}$, which have the same beamwidth of $2/N$ and the same beam gain of $M^{(\rm I)}$.}
    \end{proposition}
    Using \textbf{\textit{Proposition} \ref{prop1}} and exploiting the uniformly spaced beams, we only need to consider beam sweeping in the \textit{angular subspace} $[-1, -1+2/(M^{(\rm I)})^{2}]$. As each beam has a narrow beamwidth of $2/N$, the designed codebook $  \mathbf{W}^{(\rm I)}=\{\mathbf{w}_{\ell}^{(\rm I)} |\text{ } \ell=1,2,\ldots, L^{(\rm I)}\}$ consists of $ L^{(\rm I)}=N/(M^{(\rm I)})^{2}$ codewords. Where $ \overline{\Omega}_{m}^{(\rm I)}(\ell) = -1 + \frac{2\ell-1}{N}  + \frac{2(m-1)}{(M^{(\rm I)})^2},\text{ } \forall \ell\in \{1,2,\ldots,L^{(\rm I)}\},\forall m\in\{1,2,\ldots,(M^{(\rm I)})^{2}\}$, where $ \overline{\Omega}_{m}^{(\rm I)}(\ell)$ denotes the steering angle of subarray $m$ of the $\ell$-th codeword. Substituting $\overline{\Omega}_{m}^{(\rm I)}(\ell)$ into (\ref{angleset}), we obtain the angles steered by the $\ell$-th codeword $\{-1 + (2\hat{\ell}-1)/N + 2(q-1)/(M^{(\rm I)})^{2} |\text{ }q=1,2,\ldots,Q^{(\rm I)}\}$.

     \emph{2) Beam sweeping and identification:} In Stage 1, the BS sequentially sends $L^{(\rm I)}$ training pilots by employing the codebook $\mathbf{W}^{(\rm I)}$ to scan the angular space. For each user $k$, $ (M^{(\rm I)})^{2}$ candidate angles identified as follows. Let $P^{(\rm I)}_{k} (\ell)$ denote the received signal power of user $k$ given the codeword $\mathbf{w}_{\ell}^{(\rm I)}$. Each user $k$ selects $\hat{\ell}_{k}$ that maximizes the received power, and its candidate angles are 
     \begin{align}\label{eq:1candidate}
     \hat{\mathbf{\Omega}}_{k}^{(\rm I)} \kern-0.1em = \kern-0.1em \bigg\{ \kern-0.1em -1 &+ \frac{2\hat{\ell}_{k}-1}{N} + \frac{2(q-1)}{(M^{(\rm I)})^{2}} \Bigm| q= 1,2,\ldots,Q^{(\rm I)} \bigg\},\text{ }  \forall  k\in \mathcal{K}. 
    \end{align}
    The number of pilot symbols required in Stage 1 is given by
    \begin{equation}
        T^{(\rm I)}=L^{(\rm I)}=\frac{N}{(M^{(\rm I)})^{2}}.
    \end{equation}

  \textbf{Second stage (elimination of ambiguity):} For Stage 2, We design a dense-subarray based multi-beam codebook and determine the user angles via cross-validation.

    \emph{1) Dense-subarray based codebook:} 
    Given the candidate angles in (\ref{eq:1candidate}), Stage 2 resolves the beam ambiguity with \emph{inter-angle distance} $2/(M^{(\rm I)})^{2}$. This thus motivates us to employ the dense subarray-based multi-beam codebook \cite{you2020fast} for enabling cross-validation-based beam identification. Specifically, we redivide the entire array into $M^{(\rm I \kern-0.1em I)}$ dense subarrays, each consisting of $ N_{\rm s}=N/M^{(\rm I \kern-0.1em I)}$ antennas. To resolve the angular ambiguity, the subarray beamwidth $2M^{(\rm I \kern-0.1em I)}/N$ should be no larger than the inter-angle distance $2/(M^{(\rm I)})^{2}$. Leading to $ (M^{(\rm I)})^{2}M^{(\rm I \kern-0.1em I)}\leq N$. For each subarray $m$, its steered angle $\widetilde{\Omega}_{m},\,m\in\{1,2,\ldots,M^{(\rm I \kern-0.1em I)}\}$ can be flexibly controlled, whose corresponding beam steering vector $\widetilde{\mathbf{w}}_{m}$ is given by $\widetilde{\mathbf{w}}_{m}= e^{-\jmath \pi N_{\rm s}(m-1)\widetilde{\Omega}_{m}}\left[1,e^{-\jmath \pi  \widetilde{\Omega}_{m}},\ldots,e^{-\jmath  (N_{\rm s}-1)  \widetilde{\Omega}_{m}}\right]^{T}$. Then the full array steering vector can be constructed as 
    \begin{equation}
    \label{codeword2}
        \mathbf{w}^{(\rm I \kern-0.1em I)}=\left[\widetilde{\mathbf{w}}^{T}_{1}, \widetilde{\mathbf{w}}^{T}_{2}, \ldots, \widetilde{\mathbf{w}}^{T}_{M^{(\rm I \kern-0.1em I)}} \right]^{T}.
    \end{equation}
    Each dense subarray generates one beam with beamwidth $2M^{(\rm I \kern-0.1em I)}/N$ and the beam gain of $M^{(\rm I \kern-0.1em I)}$.
    
    The designed codebook is denoted as $ \mathbf{W}^{(\rm I \kern-0.1em I)}=\{\mathbf{w}^{(\rm I \kern-0.1em I)}_{\ell}|\text{ }\ell=1,2,\ldots,L^{(\rm I \kern-0.1em I)}\}$, containing $L^{(\rm I \kern-0.1em I)}$ codewords. We can flexibly control the steering angles of each codeword, which allows us to use the \textit{cross-validation} criterion \cite{you2020fast} to determine the true user angle in the candidate angle set (\ref{eq:1candidate}).

    \emph{2) Beam sweeping and identification:} By exploring the uniform spacing pattern of candidate angles in (\ref{eq:1candidate}) with the inter-angle distance of $2/(M^{(\rm I)})^{2}$, we can divide the entire space into $ G=N/M^{(\rm I \kern-0.1em I)}$ subspaces. As such, each subspace has an equal angular width of $2M^{(\rm I \kern-0.1em I)}/N$, which contains at most one candidate angle. Consequently, we can efficiently identify the subspaces of different users using cross-validation. Specifically, the key idea lies in utilizing multiple beams to scan the entire spatial domain to identify an initial set of candidate subspaces for the user. Then, the resulting candidate set is subsequently partitioned into two equal-size subsets. By testing only one subset instead of exhaustively searching over all subsets, the proposed approach significantly reduces the beam training overhead \cite{you2020fast}. 

    \begin{figure}
    \centering
    \includegraphics[width=0.8\linewidth]{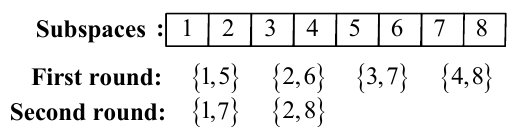}
    \caption{Illustration of \textit{Example} 1.}
    \label{example}
    \vspace{-16pt}
    \end{figure}

    \underline{\textit{\textbf{Example}}} \textbf{1}. Assume the user angle lies in one of eight uniformly spaced angles, each corresponding to a subspace. Consider $N=16$ and $M^{(\rm I \kern-0.1em I)}=2$; each codeword generates two beams. In the first round, the BS transmits $4$ codewords steered towards subspaces $\{1,5\}$, $\{2,6\}$, $\{3,7\}$, and $\{4,8\}$. By comparing the received powers, assume the codeword corresponding to $\{1,5\}$ yields the maximum received power. In the second round, to find the user subspace in $\{1,5\}$, we divide it into two subsets $\{1\}$ and $\{5\}$. Instead of sweeping both subsets, we only test one of them (e.g., $\{1\}$). To further reduce the training overhead, we use two beams to simultaneously sweep subsets from different initial sets $\{1,5\}$ and $\{3,7\}$. To maximize the inter-beam distance, the BS transmits two codewords steered towards subspaces $\{1,7\}$ and $\{2,8\}$. Since only $\{1,7\}$ overlaps with $\{1,5\}$, we only examine the codeword corresponding to $\{1,7\}$. If the received power of this codeword exceeds a certain threshold, the user subspace is determined as $\{1,5\}\cap\{1,7\}=\{1\}$; otherwise it is $\{1,5\}\setminus\{1,7\}=\{5\}$.

    Similar to \cite{you2020fast}, Stage 2 involves  $1 + \log_2 M^{(\rm I \kern-0.1em I )}$  rounds of beam sweeping. We denote by $B(r, b)$ the bin that collects the subspace steered by the $b$-th codeword in the $r$-th round. Each bin contains $M^{(\rm I \kern-0.1em I)}$ subspaces. For $r=1$, we set the bins as $ B(1,b)=\{b,b+G/(M^{(\rm I \kern-0.1em I)}),\ldots,b+(M^{(\rm I \kern-0.1em I)}-1)G/(M^{(\rm I \kern-0.1em I)})\},\text{ } \forall b\in \{1,2,\ldots,G/M^{(\rm I \kern-0.1em I)}\}$ to separate the steered subspaces in each bin as much as possible to minimize the interference between beams, where $G=N/M^{(\rm I \kern-0.1em I)}$ represents the number of subspaces. Let $P_{k}^{(\rm I \kern-0.1em I)}(r,b)$ denote the received power from the $b$-th bin of the $r$-th round of beam sweeping and $ I_{k}(r)$ represent the candidate-subspace set after the $r$-th round of beam sweeping. For $r = 1$, $I_{k}(r)$ is determined as the bin that has the highest received power, i.e., $ I_{k}(1) = B\bigl(1, b_{k}^{*}\bigr)$ with $ b_{k}^{*} = \arg\max\limits_{b\in\{1,2,\dots,G/M^{(\rm I \kern-0.1em I)}\}} P^{(\rm I \kern-0.1em I)}_{k}(1,b)$. 
    
     In each subsequent round ($ r=2,3,\ldots,\log_{2}M^{(\rm I \kern-0.1em I)}$), we set $G / (2 M^{(\rm I \kern-0.1em I)})$ bins for enabling cross-validation. For the second round, we divide the initial subspace sets from bins of the first round  into two subsets of equal size as $[B(1,b)]_{1:M^{(\rm  I \kern-0.1em I)}/2}$ and $[B(1,b)]_{M^{(\rm  I \kern-0.1em I)}/2+1:M^{(\rm  I \kern-0.1em I)}}$. We set up the second stage bins by collecting two subsets from two different initial bins, i.e, $B(2,b)=\{[B(1,b)]_{1:M^{(\rm  I \kern-0.1em I)}/2},[B(1,b+G/M^{(\rm  I \kern-0.1em I)})]_{M^{(\rm  I \kern-0.1em I)}/2+1:M^{(\rm  I \kern-0.1em I)}}\}, \text{ } \forall b=\{1,2,\ldots,G/2M^{(\rm  I \kern-0.1em I)}\}$. Similarly, each subsequent round of bins is set by collecting two subsets of two different bins from the last round. Mathematically, the $b$-th bin of the $r$-th round is given by \cite{you2020fast} 
     \begin{align*} 
     B ( r , b )& = \{ [ B ( 1 , b ) ] _ { 1 : u ( r ) } , [ B ( 1 , b + D / 2 ) ] _ { u ( r ) + 1 : 2 u ( r ) } , \nn \\  &[ B ( 1 , b ) ] _ { 2 u ( r ) + 1 : 3 u ( r ) } , [ B ( 1 , b + D / 2 ) ] _ { 3 u ( r ) + 1 : 4 u ( r ) } ,  \nn\\  &\ldots , [ B ( 1 , b ) ] _ { M^{(\rm  I \kern-0.1em I)} - 2 u ( r ) + 1 : M^{(\rm  I \kern-0.1em I)} - u ( r ) } ,  \nn\\  &[ B ( 1 , b + D / 2 ) ] _ { M^{(\rm  I \kern-0.1em I)} - u ( r ) + 1 : M^{(\rm  I \kern-0.1em I)} } \} ,\text{ } \forall b \in [ 1,2 , \ldots , D / 2 ] , 
     \end{align*} 
    where $ u ( r ) = M^{(\rm I \kern-0.1em I)} / ( 2 ^ { r - 1 } )$ and $D=G/M^{(\rm  I \kern-0.1em I)}$. The received power $P_{k}^{(\rm I \kern-0.1em I)}(r,b)$ is compared with a threshold $ P_k^{(\text{th})} = \rho \max\limits_b P_k^{(\rm I \kern-0.1em I)}(1,b)$ with $\rho=1/2$ \cite{you2020fast}. The user's candidate-subspace set is updated as
        \begin{equation} \begin{aligned} 
     I_{k}(r)=&
        \begin{cases}
        I_{k}(r-1)\cap B\bigl(r,b\bigr), 
        &\text{if }P^{(\rm I \kern-0.1em I)}_{k}(r,b)\ge P_{k}^{(\mathrm{th})},\\
        I_{k}(r-1)\setminus B\bigl(r,b\bigr),
        &\text{if }P^{(\rm I \kern-0.1em I)}_{k}(r,b)< P_{k}^{(\mathrm{th})},
        \end{cases}
     \quad k\in \mathcal{K}.
    \end{aligned}
    \end{equation}
      After $\log_2 M^{(\rm I \kern-0.1em I)}$ rounds of beam sweeping, the unique subspace of user $k$ is identified as $ I_{k}(1+\log_{2}M^{(\rm I \kern-0.1em I)})=\{\hat{g}_{k}\}$, with $\hat{g}_{k}$ representing the subspace index of user $k$. Note that each subspace has a width of $2M^{(\rm I \kern-0.1em I)}/N$ and thus covers an angular interval of $[-1 +2M^{(\rm I \kern-0.1em I)}(\hat{g}_{k}-1)/N,-1 +2M^{(\rm I \kern-0.1em I)}\hat{g}_{k}/N]$. Thus the true angle of user $k$ lies in the interval corresponding to subspace $\hat{g}_{k}$
    \begin{align*}
         \mathbf{\Omega}_{k}^{(\rm I \kern-0.1em I)} \kern-0.2em=\kern-0.1em \bigg\{\Omega \Bigm| \kern-0.1em -1 +\frac{2M^{(\rm I \kern-0.1em I)}(\hat{g}_{k}-1)}{N}\leq \Omega \leq -1 +\frac{2M^{(\rm I \kern-0.1em I)}\hat{g}_{k}}{N} \bigg\}, \text{ }\forall k\in \mathcal{K}.
    \end{align*}
The number of training symbols required in Stage 2 is
    \begin{equation}
        T^{(\rm I \kern-0.1em I)}=L^{(\rm I \kern-0.1em I)}=\frac{N}{(M^{(\rm I \kern-0.1em I)})^{2}}\left(1+\frac{\log_{2}M^{(\rm I \kern-0.1em I)}}{2}\right).
    \end{equation}
The final estimated angle of user $k$ is $\hat{\Omega} _{k}=\mathbf{\Omega}_{k}^{(\rm I)} \cap \text{ } \mathbf{\Omega}_{k}^{(\rm I \kern-0.1em I)}$.

 \begin{remark}[Beam training overhead]\label{rem:overhead}
     \emph{The total beam training overhead is}
     \begin{equation*}
         T=T^{(\rm I)}+T^{(\rm I \kern-0.1em I)}=\frac{N}{(M^{(\rm I)})^{2}}+\frac{N}{(M^{(\rm I \kern-0.1em I)})^{2}}\left(1+\frac{\log_{2}M^{(\rm I \kern-0.1em I)}}{2}\right).
     \end{equation*} 
     \emph{For ease of comparison, we set the numbers of subarrays in both stages are the same (i.e., $ M^{(\rm I)}=M^{(\rm I \kern-0.1em I)}=M$). Thus, we have $ T=N(2+\log_{2}M/2)/M^{2}=N(2+\log_{2}Q/4)/Q$, where $ Q=(M^{(\rm I)})^{2}=M^{2}$ represents the maximum number of training beams. For achieving unique identification of the user angle, it is required that $ M^{3}\leq N$.}
 \end{remark}
\begin{table}[!t]
\vspace{-6pt}
\caption{Comparison of beam training, feedback overhead and computational complexity.}
\label{overheadtable}
\centering

\scriptsize                      
\setlength{\tabcolsep}{3pt}      
\renewcommand{\arraystretch}{1.1} 

\begin{tabular}{|c|c|c|c|}
    \hline
    Schemes  &\makecell{Beam training\\ overhead}  
             &\makecell{Feedback \\overhead}  
             &\makecell{Computational \\complexity} \\
    \hline
    \makecell{Single-beam training \\ with extensive search}  
        & $N$  & $1$  & $O(N)$ \\
    \hline
    \makecell{Multi-beam training based \\on dense subarrays }  
        & $\frac{N}{ Q } \left(1 + \frac{\log_2 Q}{2}\right)$  
        & $1$  
        & $O\!\left(\frac{N\log_{2}Q}{Q}\right)$ \\
    \hline
    \makecell{Multi-beam training based \\on antenna sparse-activation}  
        & $\frac{N}{Q}+KQ$  
        & $2$  
        & $O\!\left(\frac{N+Q^{2}}{Q}\right)$ \\
    \hline
    \makecell{2-tier hierarchical search}  
        & $\min\limits_{B\leq Q} \tfrac{N}{B}+KB$ 
        & $2$  
        & $\min\limits_{B\leq Q}O\!\left(\frac{N+B^{2}}{B}\right)$ \\
    \hline
    \makecell{Proposed method}  
        & $\frac{N}{Q}\!\left(2+\frac{\log_{2}Q}{4} \right)$ 
        & $1$  
        & $O\!\left(\frac{N\log_{2}Q}{Q}\right)$ \\
    \hline
\end{tabular}
\vspace{-6pt}
\end{table}

Table~\ref{overheadtable} summarizes the beam training overhead, feedback overhead, and computational complexity of the considered schemes. When $Q$ increases, the proposed method substantially reduces the beam training overhead and computational complexity compared to existing schemes, while requiring only one feedback for each user, which is the same as the single-beam exhaustive search.

\vspace{-6pt}
 \section{Numerical Results}
\vspace{-3pt}

In this section, numerical results are presented to evaluate the performance of the proposed two-stage beam training method where the system operates at $30$ GHz. The BS is equipped with a $256$-antenna array, and $5$ users are randomly distributed on a circle around the BS with a distance of $r=50 \text{ m}$. We consider the Rician fading channel model, where the Rician factor is set to $\xi = 5 \, \text{~dB}$. The reference channel power gain is $\beta = -60 \, \text{~dB}$, with a path loss exponent $\alpha = 2.2$. We define the (reference) SNR as $\mathrm{SNR} = \frac{P\beta }{r^{\alpha} \sigma^2}$ (i.e. single-antenna-wise received SNR), where $P$ is the average transmit power and $\sigma^2$ is the average noise power. The success identification rate is given by $R_{\rm s} = \frac{1}{K} \sum_{k=1}^K \mathbb{I}(\hat{n}_k = n^*_k)$, where $\mathbb{I}(\cdot)$ is the indicator function, $\hat{n}_k$ is the best beam index based on estimated user angle $\hat{\Omega}_{k}$, and $n^*_k$ is the (true) optimal beam index for user $k$. The normalized average beam gain is expressed as $\overline{G} = \frac{1}{K} \sum_{k=1}^K \frac{1}{N} |\mathbf{v}^H(\Omega_k, N) \mathbf{u}_k|^2$.

\begin{figure}[t]
  \centering
  \begin{minipage}{0.5\linewidth}
    \subfloat[Normalized average beam gain versus reference SNR with $Q=4$.]{%
      \includegraphics[width=\linewidth]{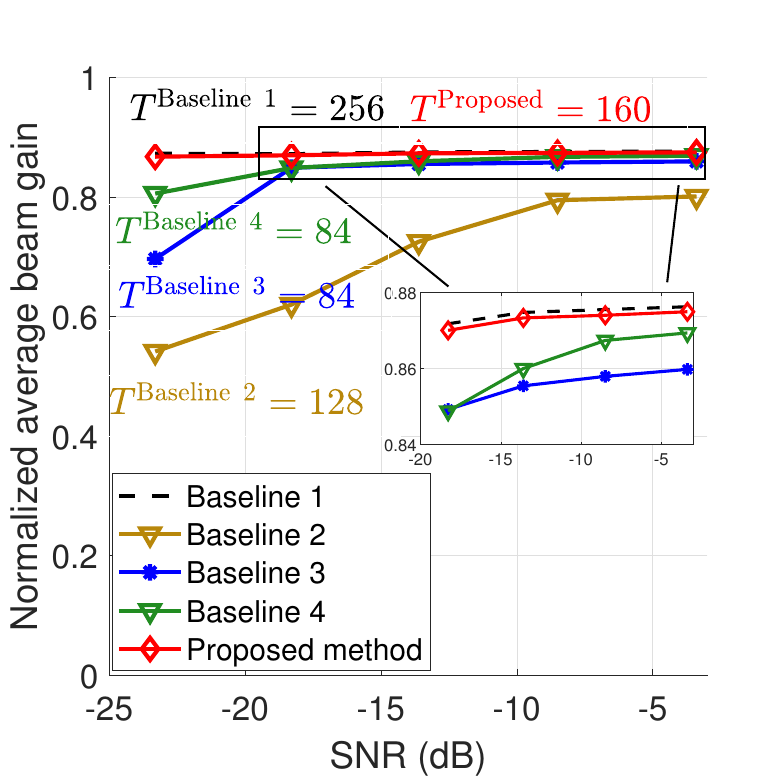}%
      \label{fig:3.1}%
    }
  \end{minipage}%
  \hfill
  \begin{minipage}{0.5\linewidth}
    \subfloat[Normalized average beam gain versus reference SNR with $Q=16$.]{%
      \includegraphics[width=\linewidth]{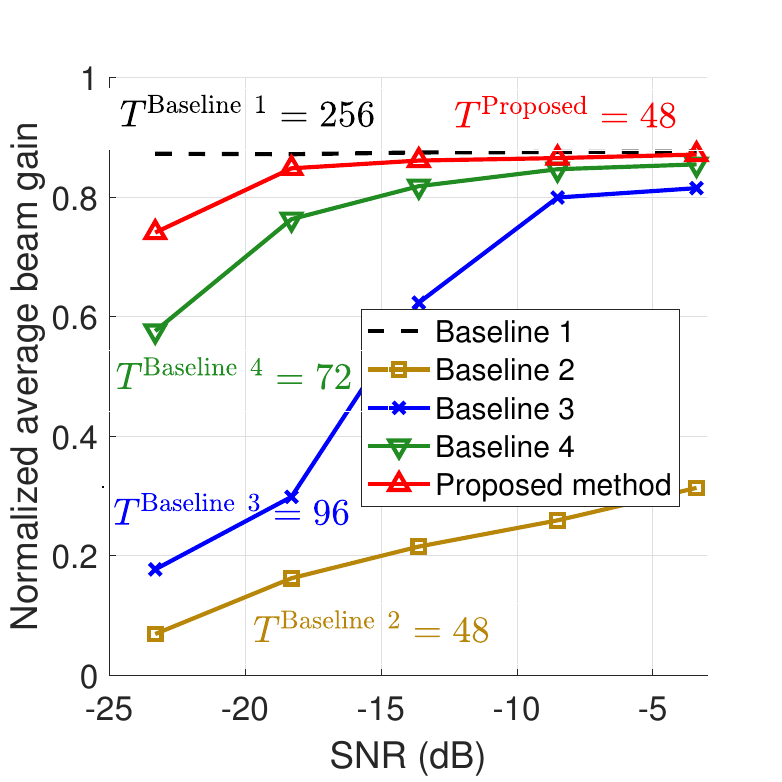}%
      \label{fig:3.2}%
    }
  \end{minipage}
  \caption{Performance comparison of the proposed two-stage multi-beam training with baselines 1-4, given different numbers of beams.}
  \label{fig:sidebyside}
  \vspace{-16pt}
\end{figure}

The numerical results are averaged over $5000$ Rician channel realizations. Three metrics are evaluated: beam training overhead, success beam-identification rate, and normalized average beam gain. All schemes use the same number of training beams $Q$ for fair comparison. The following benchmark schemes are considered for comparison. 1) Baseline 1: single-beam training with exhaustive search; 2) Baseline 2: multi-beam training based on dense subarrays \cite{you2020fast}; 3) Baseline 3: multi-beam training based on antenna sparse-activation \cite{zhou_multi-beam_2024}; and 4) Baseline 4: 2-tier hierarchical search \cite{Hierarchical_Codebook_2020}.

Figs. 4(\ref{fig:3.1}) and 4(\ref{fig:3.2}) show the normalized average beam gain versus SNR for different $Q$. Several key observations are made as follows. First, for $ Q=4$, baselines 2 and 3 suffer severe performance degradation in the low-SNR regime ($\text{SNR} <-20\,\text{~dB}$). In contrast, although the proposed method incurs higher training overhead than other multi-beam training schemes, it achieves a much higher normalized average beam gain in the low-SNR regime (e.g., $\overline{G}>0.8$ at $\text{SNR}=-23.3\, \text{~dB}$), even for the conventional 2-tier hierarchical beam search. This is because hierarchical search is vulnerable to error propagation, whereas the proposed method mitigates such effects by exploiting cross-validation, thereby enhancing robustness. When $Q=16$, power division among multiple beams leads to degraded beam gain across all multi-beam training schemes compared to the case of $Q = 4$, particularly in the low-SNR scenario. However, the proposed method significantly reduces training overhead (e.g., about $81.3\%$ reduction compared to baseline 1), while retaining a comparable normalized beam gain (approximately $0.87$) at high SNR.

\begin{figure}[t]
    \centering
    \includegraphics[width=8cm]{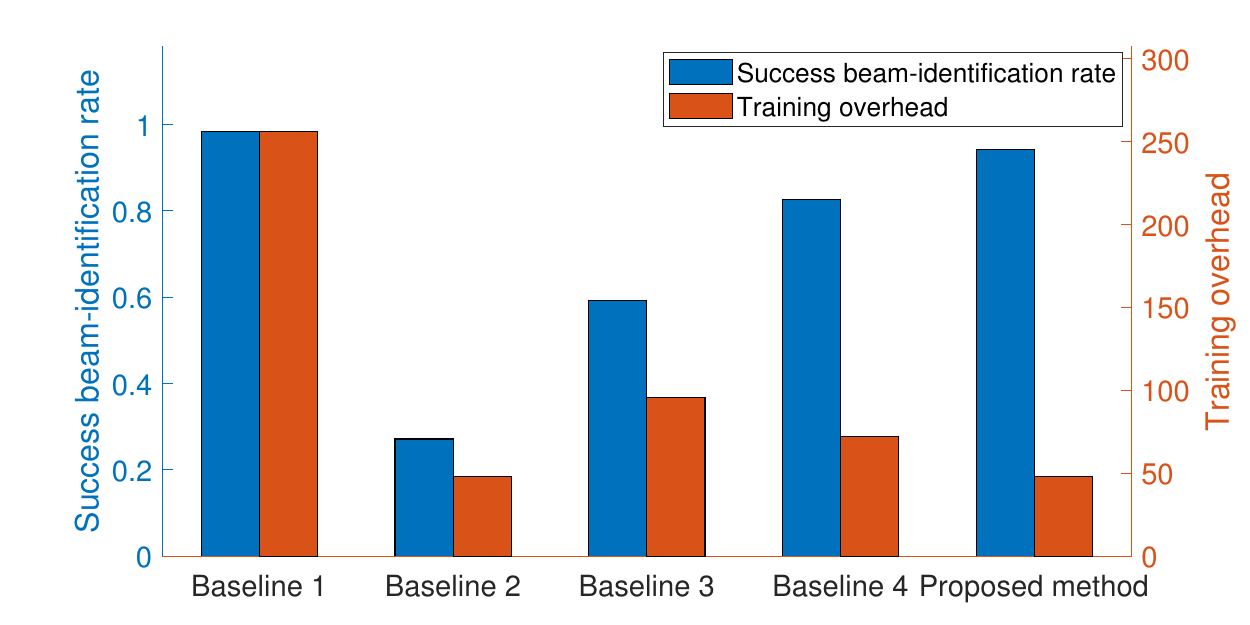}
    \caption{Comparison of success beam-identification rate and beam training overhead with reference $\text{SNR}=-18.3~\rm dB$ and $Q=16$.}
    \label{fig:overall}
    \vspace{-10pt}
\end{figure}

Fig. 5 compares the success beam-identification rate and beam training overhead of different schemes given $Q=16$ and reference $\text{SNR} = -18.3 \, \text{~dB}$. The proposed method demonstrates a higher success beam-identification rate than Baselines 2-4, given the same or even lower beam training overhead. Notably, for multi-beam training schemes, the transmit power needs to be split across multiple beams. This results in a reduced effective received SNR per beam and thus degraded success beam-identification rate, as compared to single-beam training (Baseline 1).

\vspace{-6pt}
\section{Conclusion}
\vspace{-3pt}
In this letter, we proposed a new two-stage multi-beam training method for multiuser millimeter-wave communications. This method achieves full angular resolution and flexible beam steering by dividing the antenna array into sparse and dense subarrays and designing distinct multi-beam patterns across two stages. Our proposed method achieves a high success beam-identification rate while significantly reducing beam training overhead. Notably, the proposed method requires only one feedback from users, which is the same as the conventional single-beam training.

\vspace{-6pt}
\begingroup
  \setlength{\itemsep}{-2pt}   
  \setlength{\parskip}{-6pt}   

\endgroup

\end{document}